\documentclass[prl,twocolumn,showpacs]{revtex4}
\usepackage{graphicx}
\usepackage{amsmath}
\usepackage{multirow}
\begin{document}
\bibliographystyle{apsrev} \title{Many body effects in the excitation spectrum
  of a defect in SiC} \author{Michel Bockstedte$^{1,2}$} \author{Andrea
  Marini$^{3}$} \author{Oleg Pankratov$^{2}$} \author{Angel Rubio$^{1}$}
\affiliation{$^1$ Nano-Bio Spectroscopy group and ETSF Scientific Development
  Centre, Dpto.\ F\'\i sica de Materiales, Universidad del Pa\'\i s Vasco,
  Centro de F\'\i sica de Materiales CSIC-UPV/EHU-MPC and DIPC, Av.\ Tolosa 72,
  E-20018 San Sebasti\'an, Spain.}  \affiliation{$^{2}$ Lehrstuhl
  f. Theoretische Festk\"orperphysik, Universit\"at Erlangen-N\"urnberg,
  Staudtstr. 7\,B2, D-91058 Erlangen, Germany} \affiliation{$^{3}$ European
  Theoretical Spectroscopy Facility (ETSF), Dipartimento di Fisica,
  Universit\'a di Roma Tor Vergata, I-0133 Roma, Italy} \date{\today}
\begin{abstract}
  We show that electron correlations control the photophysics of
  defects in SiC through both renormalization of the quasiparticle
  bandstructure and exciton effects. We consider the carbon vacancy,
  which is a well-identified defect with two possible excitation
  channels that involve conduction and valence band
  states. Corrections to the Kohn-Sham ionization levels are found to
  strongly depend on the occupation of the defect state. Excitonic
  effects introduce a red shift of 0.23 eV. The analysis unambigiously
  re-assigns excitation mechanism at the thresholds in photo-induced
  paramagnetic resonance measurements [J.~Dashdorj \emph{et~al.},
  J.~Appl.~Phys.~\textbf{104}, 113707 (2008)].
\end{abstract}
 \maketitle 

Optical spectra of semiconductors not only contain the excitations of
the perfect bulk crystal but also that of its omnipresent
imperfections, such as point defects. Photoluminiscence plays a
fundamental role in the experimental characterization and
identification of defects~\cite{pl}.  Despite the success of density
functional theory (DFT) in explaining the physics of defects in
covalent semiconductors~\cite{generaldft}, this common approach has
severe shortcomings: the description of the bonding of the bulk
crystal may be insufficient within common approximations for exchange
and correlation, like the local spin density approximation (LSDA), the
position of defect levels in the band gap is affected by the
well-known Kohn-Sham band gap error, and excitations cannot be
assessed rigorously. Many-body perturbation theory allows to resolve
these issues \cite{onida:02}, however, applications to defects are
scarce~\cite{defectGW} due to computational complexity.  The
quantitative prediction of absorption spectra of a wide-class of
systems from insulators to surfaces, to nanotubes and polymers became
tractable only recently by solving the Bethe-Salpeter equations (BSE)
and accounting for the G$_0$W$_0$ self-energy
~\cite{onida:02,generalbse}.

In electron paramagnetic resonance (EPR) experiments under
illumination \cite{zvanut:02,son:02,dashdorj:08}, the positively
charged carbon vacancy (V$_\text{C}^{+}$) in the 4H polytype of
silicon carbide (4H-SiC) was extensively investigated and excitation
thresholds were assigned to the defect ionization levels with respect
to the neutral defect.  The identification of this defect as an
EPR-center was established on the basis of the DFT-LSDA ground state
calculations \cite{bockstedte:03a,umeda:04}. Yet, for the accurate
interpretation of the photo-EPR data one has to consider excitation
transitions taking into account at least two competing excitation
channels: transitions to the neutral state (V$_{\text{C}}^{0}$) as
well as to the doubly positive defect ($\text{V}_\text{C}^{2+}$).
To assess the values of the ionization levels one has to evaluate
the contribution of excitonic effects that is unknown so far.
Knowledge of the ionization levels is pivotal for an understanding of
the carrier compensation in high purity semi-insulating SiC
\cite{aavikko:06}, in which vacancies are abundant compensation
centers, and for the dopant diffusion and defect kinetics
\cite{bockstedte:04}.

Defect excitations combine the many-body effects in both extended and localized
states.  The insertion of electrons into defect states invokes on-site
correlation due to electron-electron repulsion expressed by the
Hubbard $U$.  Defect-to-band transitions involve two particles, the
electron and the hole (one in the localized state and the other in the
extended state). To self-energy effects this adds excitonic
correlation energy of yet unknown size.  

The aim of this letter is to investigate many-body effects in the
absorption spectrum of a well-identified defect: a carbon vacancy in
4H-SiC. We show that those corrections strongly depend on the
occupation of the defect levels, indicating a Hubbard-$U$ which is
substantially larger than predicted by DFT-LSDA (1.09 vs.\ 0.79\,eV).
Furthermore excitonic effects introduce a sizeable red shift ($\sim
0.23$\,eV) varying with the defect charge state. More importantly, the
calculated transition energy thresholds for the two competing channels
indicate that the current interpretation of photo-EPR
spectra~\cite{zvanut:02,son:02,dashdorj:08} based on the assumption of
a photoionization via the neutral defect has to be corrected.
According to our calculations the ionization to
$\text{V}_\text{C}^{2+}$ is the only excitation mechanism at photon
energies below 2.3 eV that is responsible for the experimental
findings.

\begin{figure}
\includegraphics[width=0.9\linewidth]{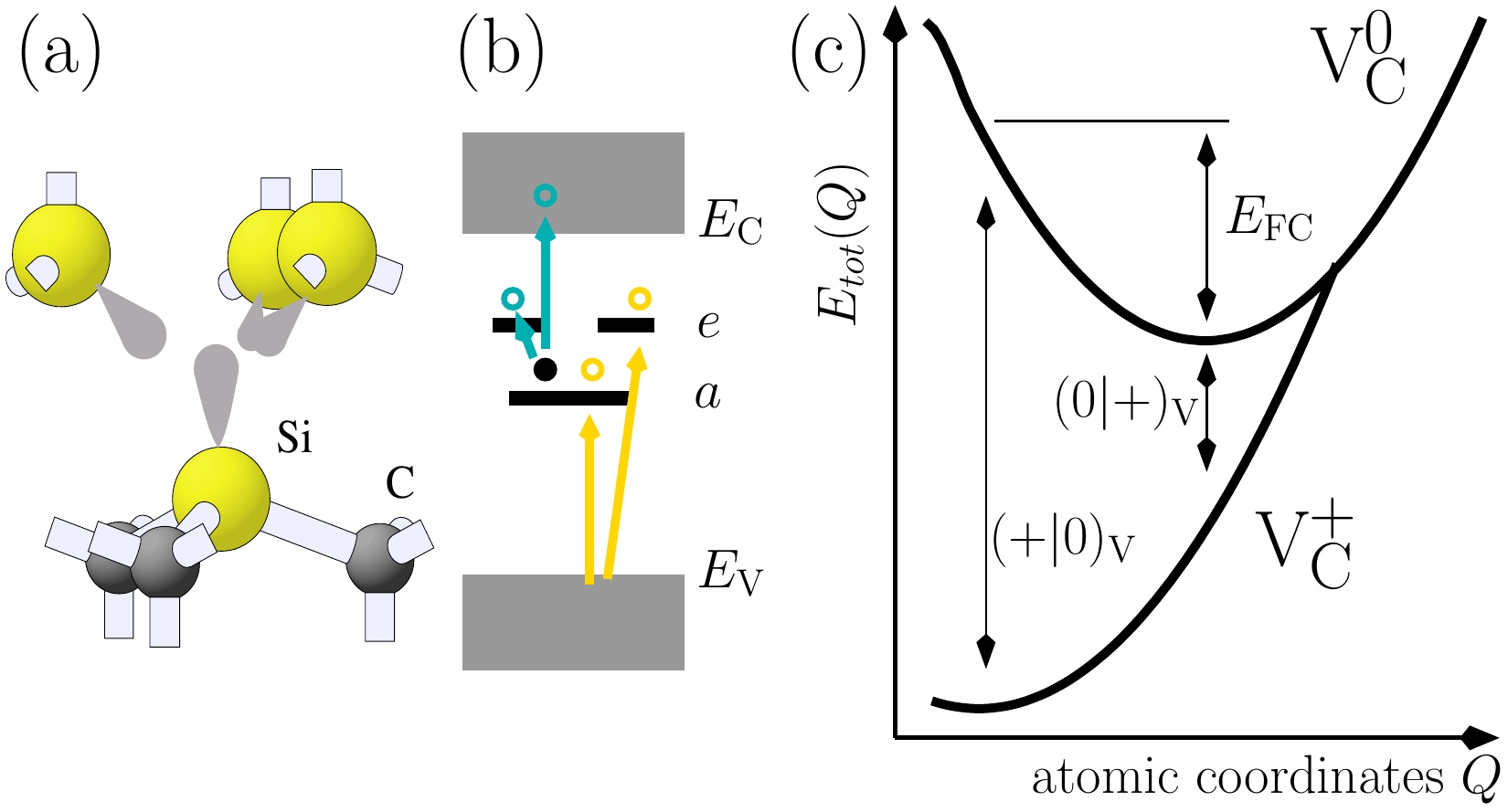}
\caption{\label{fig:vc}Carbon vacancy $\text{V}_\text{C}^+$ in 4H-SiC
(color on-line): (a) geometry -- the spin density is indicated, (b) defect levels within the band gap. In the
ground state, the $a$-level is occupied with the paramagnetic
electron. Arrows indicate the transitions from the valence band (V)
(in orange) and to the conduction band (C) (in green). (c) sketch of
the total energy $E_{\text{tot}}(Q)$ of V$_\text{C}^+$ and V$_\text{C}^0$
vs. the atomic coordinates; the QP-energies
$(+|0)_\text{V}$ and $(0|+)_\text{V}$ and the Franck-Condon shift
$E_\text{FC}$ of V$_\text{C}^0$ are indicated (cf. text).
}
\end{figure}


The defect absorption spectra and quasiparticle (QP) energies were
investigated with the YAMBO-code~\cite{marini:09}. The defect in 4H-SiC was
represented by a supercell with 288 lattice sites
\cite{endnoteDFT-LSDA}. Among the two inequivalent vacancy sites in
4H-SiC~\cite{bockstedte:03a} we focus on the experimentally investigated cubic
site.  In the G$_0$W$_0$ and BSE calculations spin was taken into account
\cite{endnoteGW}. First we turn to the fundamental band gap of 3C and
4H-SiC. The QP-corrections using the DFT-lattice constant amount to 0.91\,eV
and 0.93\,eV respectively, in agreement with earlier
$G_{0}W_{0}$-calculations~\cite{aulbur:03}. Thereby, for the gap of 4H-SiC, we
obtain $3.11$\,eV (Exp.~ 3.265\,eV~\cite{landolt}).

\begin{table}
  \caption{\label{tab:qp} QP-energies $(q|q+1)_{\text{V}}$ and
    $(q|q-1)_\text{V}$, Franck-Condon shifts
    $E_{\text{FC}}$ and thermal ionization levels $(q|q-1)$ for the
    relevant charge states of V$_{\text{C}}$ in eV. Values are listed
    as obtained within G$_0$W$_0$ and DFT.  First two columns: initial
    charge state with the occupation of
    its defect levels. $E_\text{FC}$ is taken
    from DFT calculation after electron insertion (cf. text).}

  \begin{ruledtabular}
   \begin{tabular}{lccccccccccccc}
          && \multicolumn{2}{c}{$\left(q|q+1\right)_\text{V}$}&&\multicolumn{2}{c}{$\left(q|q-1\right)_\text{V}$}&$E_{\text{FC}}$&\multicolumn{2}{c}{$\left(q|q-1\right)$}\\

                     &           &G$_0$W$_0$&DFT &&G$_0$W$_0$&DFT&&G$_0$W$_0$&DFT\\
V$_{{\text{C}}}^{2+}$&$a^0\/e^0$&           &    &&1.83&1.85&0.29&1.54&1.56\\
V$_{{\text{C}}}^{+}$&$a^1\/e^0$& 1.27       &1.24&&2.36&2.03&0.83&1.53&1.20\\
V$_{{\text{C}}}^{0}$&$a^2\/e^0$& 0.88       &0.72&&2.80&2.28&0.21&2.59&2.07\\
V$_{{\text{C}}}^{-}$&$a^2\/e^1$& 1.76       &1.39&&2.65&2.16&0.41&2.24&1.76\\
   \end{tabular}
  \end{ruledtabular}
\end{table}

Optical absorbtion, as indicated in Fig.~\ref{fig:vc}, involve the
defect levels in the band gap, a non-degenerate ($a$-level) and a
two-fold degenerated level ($e$-level). A (pseudo) Jahn-Teller
effect~\cite{bockstedte:03a} lifts the degeneracy for all charge
states except for V$_{\text{C}}^{2+}$, however, for V$_{\text{C}}^{+}$
the effect is neglegible. Electronic transitions occur predominantly
at the fixed geometry of the initial state (Franck-Condon
principle). The excitation threshold for transitions between the
valence band and the defect states is determined by the QP-energies of
the lowest unoccupied defect state (LUMO)
$\varepsilon^{{\text{QP}}}_{\text{LUMO}}$ and the valence band edge
$E_{\text{V}}$ and by the electron-hole interaction $E_{\text{e\/h}}$,
i.e. $\varepsilon^{{\text{QP}}}_{\text{LUMO}}-E_{\text{V}}-E_{\text{e\/h}}$.
For transitions from the highest occupied defect state (HOMO) to the
conduction band with the band edge $E_\text{C}$ the threshold is
$E_{\text{C}}-\varepsilon^{{\text{QP}}}_{\text{HOMO}}-E_{\text{e\/h}}$.

\paragraph{Ionization} -- We now discuss the QP-contribution to the threshold
in comparision with the DFT values to discern the many-body corrections
within the G$_{0}$W$_{0}$.  The addition of an electron to the LUMO of the
defect in the charge state $q$ (V$_{\text{C}}^{q}$) yields
V$_{\text{C}}^{q-1}$. Using $E_{\text{V}}$ as an energy reference for the
electron, the energy required for the addition is described by the QP-energy
difference $\varepsilon^{{\text{QP}}}_{\text{LUMO}}-E_{\text{V}}$, which we
denote $\left(q|q-1\right)_{\text{V}}$.  Correspondingly, the removal of an
electron from the HOMO of V$_{\text{C}}^{q}$, which yields
V$_{\text{C}}^{q+1}$, is associated with the energy gain
$\varepsilon^{{\text{QP}}}_{\text{HOMO}}-E_{\text{V}}$ denoted by
$\left(q|q+1\right)_{\text{V}}$. In DFT, the values are obtained from the
total energies.  The results are listed in Table~\ref{tab:qp}.  Consider first
$\text{V}_{\text{C}}^{2+}$ and $\text{V}_{\text{C}}^{+}$: for the addition of
an electron to the unoccupied $a$-level of $\text{V}_{\text{C}}^{2+}$, $(2+|+)_{\text{V}}$ obtained within G$_0$W$_0$ and
DFT agree to within 20\,meV. Also, for the electron removal from the
$a$-level of $\text{V}_{\text{C}}^{+}$, $(+|2+)_{\text{V}}$ obtained with G$_0$W$_0$ matches the DFT-result.  Hence,
the addition (removal) of an unpaired electron to (from) the $a$-level is
already well described within DFT.  Here the QP-correction to the
Kohn-Sham level of 0.2\,eV (insertion) and -0.23\,eV (removal) reflects the
screening by valence band electrons and defect resonances. However, for adding
a second electron to the $a$-level exchange and correlation effects beyond the
description of the DFT considerably change the mutual repulsion of the
two electrons: The G$_0$W$_0$ value for $(+|0)_{\text{V}}$ is by 0.30\,eV
larger than the DFT result.  The effective Hubbard interaction
$U=(+|0)_\text{V}-(+|2+)_\text{V}$ (cf.~\cite{zywietz:99}) among the two
electrons in the QP-treatment amounts to $U=1.09$\,eV instead of the DFT
value $U=0.79$\,eV.  For the remaining ionization levels, similar changes in
the electron-electron repulsion due to the inclusion of exact exchange and the
screended Coulomb-interaction in G$_{0}$W$_{0}$ beyond the DFT
description are found.

The thermal ionization level $\left(q|q-1\right)$ determines in
thermal equilibrium the positions of the Fermi level at which the
charge state changes from $q$ to $q-1$. The level $\left(q|q-1\right)$
differs from $\left(q|q-1\right)_V$ by the Franck-Condon shift, which
is the energy gain due to the relaxation of the vacancy after the
electron insertion (cf.\ Fig.\ \ref{fig:vc}c). The relaxation leads to
an effective electron-electron repulsion $U_\text{eff}$ smaller than
$U$. To assess $\left(q|q-1\right)$ within the G$_0$W$_0$-approach, we
approximate the Franck-Condon shift $E_\text{FC}$ by the DFT
values. The results are listed in Table~\ref{tab:qp}. For
V$_\text{C}^{+}$, DFT predicts a negative value for
$U_\text{eff}$ \cite{zywietz:99}. Within DFT we obtain
$U_\text{eff}=-0.36$\,eV. This would imply that the charge state
V$_\text{C}^+$ is unstable and in contradiction to the
findings~\cite{son:02,zvanut:02} should hardly be observable by EPR in
thermal equilibrium. However, due to the QP-corrections, the levels
$(2+|+)$ and $(+|0)$ become almost equal (i.e. $U_\text{eff}\approx
0$), thus resolving the contradiction.  Similarly, a
negative-$U$ effect is found for V$_\text{C}^{-}$. Here, after the
QP-correction, $U_\text{eff}$ remains negative with a value of
-0.35\,eV.

\begin{figure}
\includegraphics[width=0.93\linewidth]{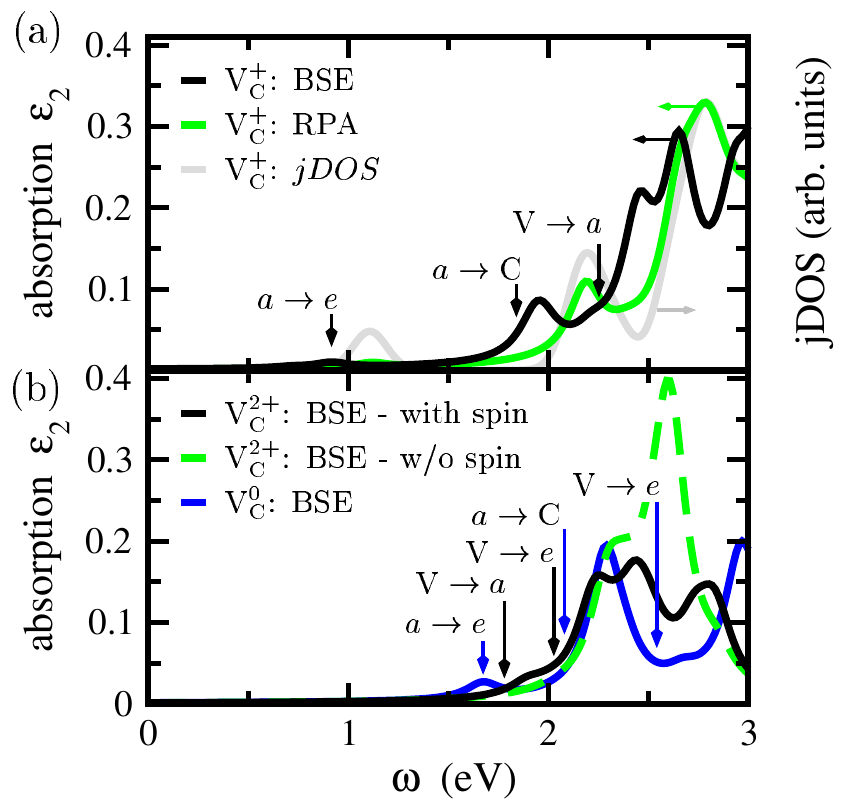}
\caption{\label{fig:vc-exc} (color online) Absorption spectra of
  $\text{V}_{\text{C}}$ within G$_{0}$W$_{0}$-BSE: (a) for
  $\text{V}_{\text{C}}^{+}$ in comparision with the G$_{0}$W$_{0}$-RPA
  spectrum and the joint densisty of states (jDOS) of the transistions with
  the $a$-level, (b) for $\text{V}_\text{C}^{0}$ and $\text{V}_\text{C}^{2+}$
  as calculated with spin-polarization and without (for the latter). The onset
  of excitation channels is indicated. A broadening of 0.1\,eV was used for
  all spectra.}
\end{figure}

\begin{table}
  \caption{\label{tab:bse}Ionization (G$_0$W$_0$) and excitation (BSE)
    thresholds as well as experimental data. Also the excitation
    energy of transition between the $a$ and $e$ level are
    given. Values are in eV.}
\begin{ruledtabular}
\begin{tabular}{lcccrcccrccc}
                &\multicolumn{3}{c}{V$_{\text{C}}^{2+}$}&&\multicolumn{3}{c}{V$_{\text{C}}^{+}$}
                                                          &&\multicolumn{3}{c}{V$_{\text{C}}^{0}$}\\
                &G$_{0}$W$_{0}$&BSE&$ E_{\text{e\/h}}$&&G$_{0}$W$_{0}$&BSE&$ E_{\text{e\/h}}$&
&G$_{0}$W$_{0}$&BSE&$E_{\text{e\/h}}$\\
$a\rightarrow e$&    &    &    &&1.09&0.89&0.20&&1.92&1.68&0.24\\
V$\rightarrow a$&1.83&1.65&0.18&&2.39&2.16&0.23&&\\
V$\rightarrow e$&2.14&1.96&0.18&&2.36&2.13&0.23&&2.80&2.56&0.21\\
$a\rightarrow \text{C}$&   &  &  &&1.84&1.61&0.23&&2.23&2.02&0.21\\
\hline
\multicolumn{5}{l}{Experimental thresholds}\\
\multicolumn{5}{l}{quenching of V$_\text{C}^{+}$}&1.47\footnote{Ref.~\cite{son:02}}&&1.6\footnote{Ref.~\cite{dashdorj:08}}\\
\multicolumn{5}{l}{restoring of V$_\text{C}^{+}$}&1.81$^{a}$&&1.9$^{b}$\\
 \end{tabular}
\end{ruledtabular}
\end{table}

\begin{figure}
\includegraphics[width=0.93\linewidth]{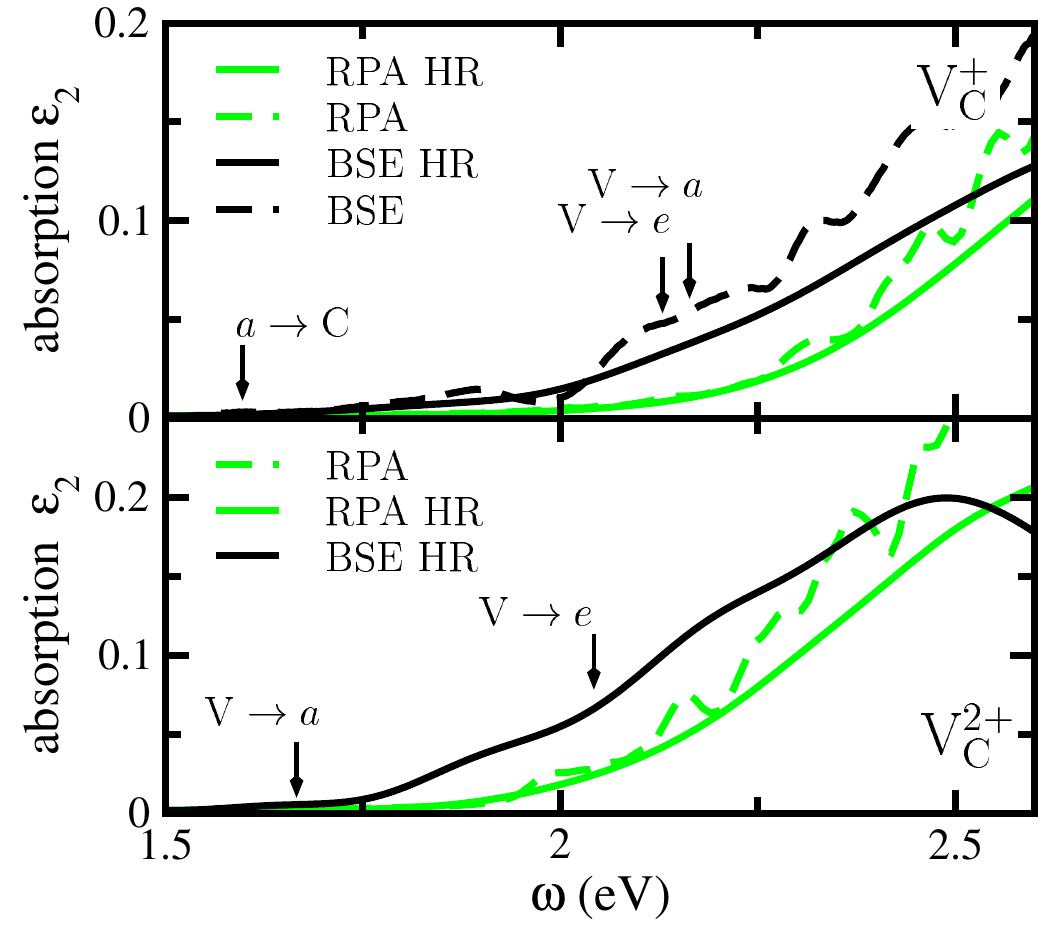}
\caption{\label{fig:rpa} (color online) Absorption spectra of
  V$_{\text{C}}^{+}$ and V$_{\text{C}}^{2+}$. G$_{0}$W$_{0}$-RPA and
  extrapolated G$_{0}$W$_{0}$-BSE spectra for high k-point
  density. The effect of Franck-Condon shifts are shown (indicated by
  the extension HR). A broadening of 0.01\,eV was used for the raw
  G$_{0}$W$_0$-RPA spectra.}
\end{figure}

\paragraph{Optical properties} -- Photo-EPR experiements~\cite{zvanut:02,son:02,dashdorj:08} were
conducted with $\text{V}_{\text{C}}^+$ as the initial or the final
state. The electronic transitions, that quench the paramagnetic state
and include the ionization of V$_\text{C}^+$ into
$\text{V}_\text{C}^{2+}$ or $\text{V}_\text{C}^{0}$, are shown in
Fig.~\ref{fig:vc}: (i) excitation from the $a$-level to the conduction
band  ($a\rightarrow C$), or (ii) from the valence band to the defect levels
$a$ or $e$ ($V\rightarrow a$ or $V\rightarrow e$). Restoring
the EPR-signal of V$_{\text{C}}^{+}$ starts from
$\text{V}_{\text{C}}^{2+}$ or $\text{V}_{\text{C}}^0$ as initial
states by exciting an electron from the valence band to the $a$-level
($\text{V}_{\text{C}}^{2+}$) or from the $a$-level to the conduction
band ($\text{V}_{\text{C}}^{0}$) and implies the subsequent
dissociation of the electron-hole pair.  We calculated absorption
spectra for $\text{V}_{\text{C}}^{2+}$, V$_{\text{C}}^{+}$ and,
$\text{V}_{\text{C}}^0$ to unravel the relevance of competing channels
and to assess the electron-hole coupling.  For the latter purpose we
compared the spectra with the inclusion of the electron-hole
interaction via the BSE and without it, in the random phase
approximation (RPA). In the investigated energy range only absorbtion via defect states
contributes to spectra due to the indirect band gap. The spectra are
shown in Fig.~\ref{fig:vc-exc} \cite{endnotedisp}. In
Table~\ref{tab:bse}, we list the ionization and excitation thresholds.
We also calculated RPA spectra, shown in Fig.~\ref{fig:rpa}, for the
denser $4\times 4\times 4$ k-point mesh~\cite{endnotedisp1} that are
converged with respect to the density of extended
states. Since such calculations are not feasible for the BSE,
corresponding BSE spectra were extrapolated by a convolution
ansatz. Spectral broadening due to non-vertical transitions is accounted for
by applying a Huang-Rhys factor for the coordinate that connects the geometry of the
ground state and the ionized state (broading amounts to
$\sim$0.26\,eV).

First we address the spectra of $\text{V}_{\text{C}}^{+}$ obtained for
the special k-point (Fig.~\ref{fig:vc-exc}a). Excitonic effects lead
to an almost rigid red shift of the RPA-spectrum by
$E_{\text{eh}}=0.23\,\text{eV}$. The same effect is found also for
$\text{V}_{\text{C}}^{2+}$, and $\text{V}_{\text{C}}^0$ albeit with
different values of $E_{\text{eh}}$ (cf.~Table~\ref{tab:bse}).  The
structure of the RPA-spectrum is dominated by band structure
effects. This is demonstrated by comparison with the joint density of
states (jDOS) of the transitions involving the $a$-level, which
essentially reproduces the RPA-spectrum. The spectral broadening by
Huang-Rys factors, however, reduces these features as seen in
Fig.~\ref{fig:rpa}.  Deviations between the BSE and RPA spectra,
besides the red-shift, stem from bright and dark exciton solutions of
the BSE due to the spin-dependent electron-hole exchange coupling.
The reduction of absorption due to dark exciton states is, however,
not obtained in a spin-averaged treatment. This is seen in
Fig.~\ref{fig:vc-exc}b, where we compare the BSE-spectra of
V$_\text{C}^{2+}$ as obtained in the two ways.

The vertical excitation $a \rightarrow e$ between the defect levels
occur at 0.89 and 1.68\,eV for V$_{\text{C}}^{+}$ and
V$_{\text{C}}^{0}$ well below the onset of transitions involving
extended states.  An important result is that the transitions to the
conduction band set in at energies (vertical
transitions at 1.61 and 2.02\,eV for V$_{\text{C}}^{+}$ and
V$_{\text{C}}^{0}$, respectively) below the threshold for excitation
from the valence band (2.13\,eV for V$_{\text{C}}^{+}$ and 2.56\,eV
for V$_{\text{C}}^{0}$). 

From recent time-resolved photo-EPR experiments~\cite{dashdorj:08}
cross sections for quenching and restoring of the EPR signal of
$\text{V}_{\text{C}}^+$ are available with thresholds at 1.6 and
1.9\,eV respectively. Steady state photo-EPR~\cite{son:02}
yields values of 1.47 and 1.81\,eV, respectively (cf.~Table
\ref{tab:bse}). The calculated onset for excitation
$a\rightarrow\text{C}$ at 1.61\,eV agrees well with the experiments.
A rise of the quenching cross section at 2.4\,eV~\cite{dashdorj:08}
should be related to our finding for the onset of the transitions
$V\rightarrow a$ at 2.16\,eV and direct ionization above 2.4\,eV that
yields V$_{\text{C}}^{0}$.  For restoring $\text{V}_{\text{C}}^{+}$,
however, the excitation of an electron-hole pair with its subsequent
thermal dissociation or direct ionization is required with thresholds
of 1.65 and 1.83\,eV, respectively.  The comparison with experiment
indicates that direct ionization is the relevant process.  The present
analysis of competing excitations suggests a re-interpretation of the
photo-EPR spectra~\cite{zvanut:02,son:02,dashdorj:08} in terms of a
photoionization of $\text{V}_\text{C}^{+}$ via
$\text{V}_\text{C}^{2+}$ instead of $\text{V}_{\text{C}}^{0}$ as
considered there. Consequently, the ionization levels deduced from the experiments
have to be reassigned correspondingly and corrected for the exciton binding energy.

In summary we conducted GW+BSE full fledge calculations of the photophysics of
the carbon vacancy in 4H-SiC. We demonstrate the importance of adding
correlation effects at that level, in spite of the fact that the structure is
well described at the DFT-level. Charge state-dependent corrections include
the ionization levels ( up to $0.5$\,eV), the electron-electron repulsion
$U$ ($\sim 0.3$\,eV) and the electron-hole attraction ($\sim 0.23$\,eV). 
This has implications for the analysis of defects in materials, such as oxides,
where the correct despription of the bonding is an issue.
We showed that ionization of the defect into V$_{\text{C}}^{2+}$ and
V$_{\text{C}}^{0}$ compete.  However, for the range of photon energies in
which photo-EPR experiment identified thresholds only the former ionization
channel is active. Excitation spectra are red shifted by strong excitonic
effects. This unambigously rectifies the earlier experimental assignment of
transitions and the assessment of ionization levels. These are pivotal
for the defect diffusion and for unraveling the mechanisms carrier
compensation in semi-insulating SiC and hence for applications.

The authors thank Drs N.\ T.\ Son and M.\ E.\ Zvanut for fruitful
discussions. We acknowledge funding by the Deutsche
Forschungsgemeinschaft (BO1851/2-1), Spanish MEC
(FIS2007-65702-C02-01), ACI-promciona project (ACI2009-1036), and
"Grupos Consolidados UPV/EHU del Gobierno Vasco" (IT-319-07), the
European Community through e-I3 ETSF project (Contract Number 211956).
Support by the Barcelona Supercomputing Center, "Red Espanola de
Supercomputacion" is acknowledged.


\begin{thebibliography}{26}

\bibitem{pl}
 W.~J.~Choyke and L.~Patrick, Phys.~Rev.~B \textbf{4}, 1843 (1971);
 L.~W.~Song, X.~D.~Zhan, B.~W.~Benson, and G.~D.~Watkins, Phys.~Rev.~B
  \textbf{42}, 5765 (1990).

\bibitem{generaldft} A.~F.~Kohan, G.~Ceder, D.~Morgan, and C.~G.~Van de Walle,
  Phys.~Rev.~B \textbf{61}, 15019 (2000); S.~B.~Zhang and S.~H.~Wei,
  Phys.~Rev.~Lett.~\textbf{86}, 1789 (2001); J.~Neugebauer and C.~G.~Van de
  Walle, Phys.~Rev.~Lett.~\textbf{75}, 4452 (1995).

\bibitem{onida:02}
G.~Onida, L.~Reining, and A.~Rubio,
Rev.~Mod.~Phys. \textbf{74},  601 (2002).


\bibitem{defectGW}
 M.~P. Surh, H.~Chacham, and S.~G. Louie,
  Phys.~Rev.~B \textbf{51},  7464 (1995);
 Y.~Ma and M.~Rohlfing,
  Phys.~Rev.~B \textbf{77}, 115118 (2008);
P.~Rinke, A.~Janotti, M.~Scheffler,  and C.~G. Van~de Walle, 
Phys.~Rev.~Lett. \textbf{102}, 026402 (2009).

\bibitem{generalbse}
M.~Rohlfing and
S.~G.~Louie, Phys.~Rev.~Lett.~\textbf{80}, 3320 (1998); S.~Albrecht,
L.~Reining,R.~Del Sole, and G.~Onida, Phys.~Rev.~Lett.~\textbf{80}, 4510
(1998); L.~Wirtz, A.~Marini, and A.~Rubio, Phys.~Rev.~Lett.~96, 126104 (2006); 
D.~Varsano, A.~Marini, and A.~Rubio, Phys.~Rev.~Lett.~\textbf{101}, 133002 (2008).
L. Yang, M.L. Cohen, S.G. Louie, Phys. Rev. Lett. \textbf{101}, 186401 (2008).
\bibitem{zvanut:02}
 M.~E. Zvanut and V.~V.~Konovalov,
  Appl.~Phys.~Lett. \textbf{80}, 410 (2002).

\bibitem{son:02}
 N.~T. Son, B.~Magnusson, and E.~Janz{\'e}n,
  Appl.~Phys.~Lett.~\textbf{81}, 3945 (2002).

\bibitem{dashdorj:08}
 J.~Dashdorj, M.~E.~Zvanut, and J.~G. Harrison, 
 J.~Appl.~Phys. \textbf{104}, 113707 (2008).


\bibitem{bockstedte:03a}
M.~Bockstedte, M.~Heid, and O.~Pankratov,
  Phys.~Rev.~B \textbf{67}, 193102 (2003).

\bibitem{umeda:04}
T.~Umeda \emph{et~al.}, 
  Phys.~Rev.~B \textbf{69},  121201(R) (2004).

\bibitem{aavikko:06}
R.~Aavikko, \emph{et~al.},
  Phys.~Rev.~B \textbf{75}, 085208 (2007).

\bibitem{bockstedte:04} Z.~Zolnai, N.~T.~Son, C.~Hallin, and
  E. Jan{\'z}en, J.~Appl.~Phys.~\textbf{96}, 2406 (2004). 

\bibitem{marini:09}
A.~Marini, C.~Hogan, M.~Gr{\"u}ning, and D.~Varsano,
  Comp.~Phys.~Comm. \textbf{180}, 1392 (2009).

\bibitem{endnoteDFT-LSDA} For the DFT-LSDA ground state, the ABINIT-package
  [X. Gonze \emph{et~al.} Computational Materials Science \textbf{25}, 478
  (2002)] with norm conserving pseudopotentials [N.~Troullier and
  J.~L. Martins, Phys.~Rev.~B \textbf{43}, 1993 (1991)], a plane wave basis
  (energy cut-off of 30\,Ry) and, to reduce the defect-defect interaction, the
  special k-point $(0,0,\frac{1}{4})$ was used.  At this k-point the highest
  valence band lies 60\,meV below the valence band edge. Ionization levels in
  Table~\ref{tab:qp} were corrected for this dispersion. Tests for cells with
  576 lattice sites confirmed DFT-results.


\bibitem{endnoteGW} $G_{0}W_{0}$-calculations employ the plasmon pole
  approximation and include local-field effects in the dielectric function
  $\varepsilon$ and self energy $\Sigma$ with an energy cut-off of
  6\,Ry. Convergence was found to be better than 0.1\,eV with respect to these
  parameters and the number of empty bands. QP-energies for V$_{\text{C}}^{+}$
  and V$_{\text{C}}^{-}$ (with an unpaired electron) and all spectra were
  evaluated including spin. We employ $\varepsilon(\omega)$ of the defect-free
  supercell as the low energy transition between the defect levels results in
  an artifical enhancement of $\varepsilon(\omega=0)$. Tests show that this
  leads to an underestimation of QP-corrections of about 0.1\,eV. This error
  reduces with increasing cell size.


\bibitem{aulbur:03}
W. G. Aulbur, M. St{\"a}dele, and A. G{\"o}rling,
Phys.~Rev.B \textbf{62}, 7121 (2000).

\bibitem{landolt}
  \emph{Semiconductors Physics of Group IV Elements and III-V
  Compounds}, edited by K.~Hellwege and O.~Madelung 
  \emph{Landolt-B\"ornstein, New Series} (Springer, Berlin).

\bibitem{zywietz:99} A.\ Zywietz, J. Furthm\"uller, and F. Bechstedt, Phys.\ Rev.\ B \textbf{59}, 15166 (1999).

\bibitem{endnotedisp} Spectra in Fig.~\ref{fig:vc-exc} 
only include bands at the special k-point. The indicated onset of transitions
reflects this.

\bibitem{endnotedisp1}
To avoid artefacts from the dispersion of defect levels, the values at the
special k-point were used at all k-points.

\end{thebibliography}
\end{document}